\documentclass[pre,twocolumn,10pt,aps,amsmath,amssymb,amsfonts,english,floatfix]{revtex4-1}
\pdfoutput=1

\usepackage{graphicx}
\usepackage[table]{xcolor}

\usepackage{hyperref}

\usepackage{color}

\newcommand{\refeq}[1]{Eq.~(\ref{#1})}

\newcommand{\refsec}[1]{Section~\ref{#1}}
\newcommand{\refapp}[1]{Appendix~\ref{#1}}
\newcommand{\reftaband}[2]{Tables~\ref{#1} and \ref{#2}}
\newcommand{\reffig}[1]{Figure~\ref{#1}}
\newcommand{\reffigand}[2]{Figures~\ref{#1} and \ref{#2}}
\newcommand{\refcite}[1]{Ref.~\cite{#1}}
\newcommand{\punc}[1]{\,\text{#1}}
\newcommand{\sub}[1]{_{\text{#1}}}

\newcommand{\etal}{{\it et al.}}

\newcommand{\scO}{\mathcal{O}}
\newcommand{\dsZ}{\mathbb{Z}}
\newcommand{\scR}{\mathcal{R}}
\newcommand{\goS}{\mathfrak{S}}

\newcommand{\ket}[1]{\lvert #1 \rangle}

\DeclareMathOperator{\Tr}{Tr}

\begin{document}

\title{Eigenstate thermalization hypothesis in quantum dimer models}

\author{Zhihao Lan and Stephen Powell}
\affiliation{Centre for the Mathematics and Theoretical Physics of Quantum Non-equilibrium Systems} 
\affiliation{School of Physics and Astronomy, University of Nottingham, Nottingham NG7 2RD, United Kingdom}

\begin{abstract}
We use exact diagonalization to study the eigenstate thermalization hypothesis (ETH) in the quantum dimer model on the square and triangular lattices. Due to the nonergodicity of the local plaquette-flip dynamics, the Hilbert space, which consists of highly constrained close-packed dimer configurations, splits into sectors characterized by topological invariants. We show that this has important consequences for ETH: We find that ETH is clearly satisfied only when each topological sector is treated separately, and only for moderate ratios of the potential and kinetic terms in the Hamiltonian. By contrast, when the spectrum is treated as a whole, ETH breaks down on the square lattice, and apparently also on the triangular lattice. These results demonstrate that quantum dimer models have interesting thermalization dynamics.
\end{abstract}

\date{\today}

\maketitle

\section{Introduction}
\label{SecIntroduction}

Considerable attention has recently been devoted to the question of whether and how an isolated quantum many-body system thermalizes \cite{review1, review2, review3}. At the center of the topic is the eigenstate thermalization hypothesis (ETH), which states that each energy eigenstate of a generic many-body Hamiltonian is indistinguishable from a microcanonical ensemble with the same energy \cite{Srednicki1999} (see \refsec{SecBackground} for a more precise statement). ETH is expected to be valid for nonintegrable systems when energy is the only conserved quantity, but is known to fail for integrable systems \cite{integrable_review}. In the latter case, the dynamics can be described by the generalized Gibbs ensemble (GGE), obtained by maximizing the entropy subject to the appropriate mean values of all conserved quantities. Both ETH and the GGE have recently been studied in experiments with ultracold atomic gases \cite{exp1,exp2}.

In this work we use exact diagonalization to study ETH in the quantum dimer model (QDM) \cite{QDM_review}. The QDM was originally introduced by Rokhsar and Kivelson (RK) as an effective description of quantum antiferromagnets \cite{QDM_RK}, and can potentially be simulated using cold atoms \cite{Rydberg_QDM,cold_QDM}. Its ground-state properties have been studied extensively, and it is known to exhibit liquid phases with topological order on nonbipartite lattices in two dimensions (2D) \cite{QDM_liquid1, QDM_liquid2}.

Our study of the 2D QDM combines a number of features that have proven to be of interest in other recent work on ETH \cite{2D_ETH, 2D_ETH_off, 2D_TFIM, constrained_anyon, UedaPRE16}: First, it continues the progress of ETH studies from the familiar territory of 1D systems \cite{1D_ETH1,1D_ETH2,1D_ETH3,1D_ETH4} into higher dimensions. Questions about ETH in systems in two or more dimensions, such as the transverse-field Ising model on the square lattice \cite{2D_ETH, 2D_ETH_off, 2D_TFIM}, are of great interest but challenging due to the rapid increase of the Hilbert-space dimension with system size. Second, the Hilbert space of the QDM is spanned by dimer configurations subject to strong local constraints. The study of ETH in constrained systems has recently been initiated by Chandran \etal\ \cite{constrained_anyon}, who considered non-Abelian anyon chains, motivated by the question of whether the constraints can hinder thermalization.

A third interesting feature of the QDM is that global constraints on the local plaquette-flip dynamics cause the Hilbert space to split into topological sectors, characterized by a pair (in 2D) of winding numbers. A crucial distinction can be drawn between lattices that are bipartite, such as square, and those that are not, such as triangular: For bipartite lattices, in which the sites can be divided into two sublattices such that all nearest-neighbor pairs are on opposite sublattices, the winding numbers are integers and the number of sectors grows with size \cite{Square_QDM}. By contrast, for nonbipartite lattices the winding numbers can be defined only modulo \(2\), and there are four sectors defined by a pair of \(\dsZ_2\) invariants \cite{Triangular_QDM1, Triangular_QDM2}. Noting the effect of topological blocking on quench dynamics \cite{Topo_blocking}, and recalling the case of integrable models, where the extensive number of conserved quantities invalidates ETH, it is natural to ask whether the existence of these topological invariants will influence ETH in the QDM \cite{UedaPRE16}.

Our main finding is that the QDM obeys ETH with certain important caveats: On both the square and triangular lattices, ETH is only clearly satisfied when each topological sector is treated separately. If the spectrum is treated as a whole, ETH apparently breaks down, at least on the square lattice, as a result of separation of the spectra into bands corresponding to different topological sectors. On the triangular lattice, our results are less clear-cut, because of limitations on the accessible system geometries, and it is possible that ETH is restored at larger system sizes, even if the spectrum is treated as a whole. We also find that ETH breaks down when the ratio of the potential- and kinetic-energy terms in the Hamiltonian becomes large, which we understand as the approach to the trivially integrable point when this ratio is infinite. 

In \refsec{SecBackground}, we give a brief definition of ETH and introduce the QDM on both the square and triangular lattices. We also discuss the distinction between the topological sectors of the square and triangular lattices and outline the main aims of the paper. We present our main results in \refsec{SecResults} and conclude in \refsec{SecConclusions}.

\section{Background and aims}
\label{SecBackground}

\subsection{Eigenstate thermalization hypothesis}
\label{SecETH}

ETH amounts to a conjecture regarding the statistical distribution of the matrix elements of any local observable $\scO$ in the eigenbasis of the Hamiltonian \(H\) for large system size \(N\). For eigenstates \(\lvert m \rangle\) and \(\lvert n \rangle\), with eigenvalues \(E_m\) and \(E_n\), it states that \cite{Srednicki1999}
\begin{equation}
\langle m|\mathcal{O}|n\rangle = d_{\mathcal{O}}(\bar{E})\delta_{mn}+\frac{r_{mn}}{\sqrt{e^{S(\bar{E})}}}f_{\mathcal{O}}(\bar{E},E_m-E_n)\punc,
\end{equation}
where $\bar{E}=(E_m+E_n)/2$, $\delta_{mn}$ is the Kronecker delta, and $S(E)$ is the (extensive) thermodynamic entropy at energy $E$; $r_{mn}$ are independent Gaussian random variables with zero mean and unit variance; and $d_{\mathcal{O}}$ and $f_{\mathcal{O}}$ are functions that are smooth on the scale of the many-body level spacing \(\sim \bar{E} e^{-S(\bar{E})}\). The statement is expected to apply only for eigenstates that are sufficiently far from the edges of the spectrum (and, in particular, not to the ground state or low-lying excited states).

The function \(d_{\scO}\) can be related to the expectation value of $\mathcal{O}$ in the canonical ensemble at temperature \(T = 1/\beta\),
\begin{equation}
\langle\mathcal{O}\rangle_T = \frac{\Tr \mathcal{O} e^{-\beta H}}{\Tr e^{-\beta H}}\punc{.}
\label{EqThermalEV}
\end{equation}
Replacing the sums over eigenstates by integrals over energy, which are then evaluated by the method of steepest descent \cite{Srednicki1999}, one finds
\begin{equation}
d_{\mathcal{O}}(E_T)=\langle\mathcal{O}\rangle_T + O(N^{-1})\punc{,}
\label{diag_ETH}
\end{equation}
where $E_T = \langle H \rangle_T$ is the average energy at temperature $T$.

The statement of ETH that we wish to test here is the following: the diagonal matrix elements \(\langle n|\mathcal{O}|n\rangle\) for energy \(E_n\) near \(E_T\) are distributed around the value \(d_{\scO}(E_T)\), with fluctuations that are exponentially small in \(N\), and this value furthermore approaches \(\langle\mathcal{O}\rangle_T\) as \(N \rightarrow \infty\).

\subsection{QDM on square and triangular lattices}
\label{SecModels}

The Hilbert space of the QDM is spanned by the set of basis states \(\ket{\Psi}\) for all possible close-packed dimer configurations \(\Psi\), in which each site forms a dimer with exactly one of its nearest neighbours. We use the standard RK Hamiltonian \cite{QDM_RK} with a kinetic term that flips dimers around the shortest possible loop, referred to as a plaquette, and potential energy proportional to the number of flippable plaquettes.

The Hamiltonian can be written as \cite{QDM_review}
\begin{align}
H^\alpha &= \sum_p\big({-t} T_p^\alpha + V P_p^\alpha\big)\punc{,}
\label{eqn:QDMs}
\end{align}
where $\alpha=\triangle$ and $\square$ for the triangular and square lattices respectively. The operator \(T_p^\alpha\) flips dimers around plaquette \(p\) or gives \(T_p^\alpha \ket{\Psi} = 0\) if \(p\) is not flippable in configuration \(\Psi\), while \(V_p^\alpha = (T_p^\alpha)^2\) is diagonal in the configuration basis, with matrix element \(1\) if \(p\) is flippable and \(0\) otherwise.
\begin{figure}
\begin{center}
\includegraphics{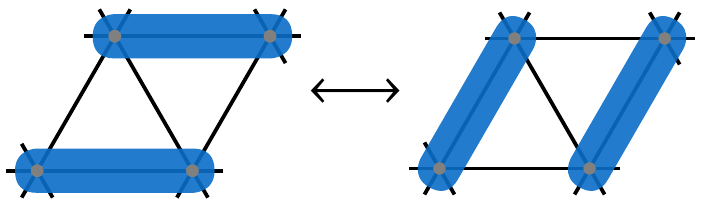}
\end{center}
\caption{Action of \(T_p^\triangle\), the plaquette-flip operator for the triangular lattice. We show only one out of the three possible orientations of the rhombus-shaped plaquette \(p\).}
\label{FigTrianglePlaquetteFlip}
\end{figure}
For the triangular lattice, the smallest possible flippable loop is a rhombus, and the action of \(T_p^\alpha\) is illustrated in \reffig{FigTrianglePlaquetteFlip}.
\begin{figure}
\begin{center}
\includegraphics{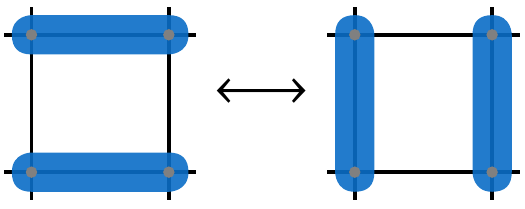}
\end{center}
\caption{Action of \(T_p^\square\), the plaquette-flip operator for the square lattice.}
\label{FigSquarePlaquetteFlip}
\end{figure}
For the square lattice, the plaquettes are squares, as shown in \reffig{FigSquarePlaquetteFlip}.

The ground state of the QDM is known exactly in certain cases: For \(V = +\infty\), the ground state is a staggered configuration in which there are no flippable plaquettes, while for $V = -\infty$, it is a columnar configuration which maximizes their number. A third special case is the RK point \(t = V\), where the exact ground state is an equal-amplitude superposition of all allowed dimer configurations \cite{QDM_RK}.

In this work, we will focus on a different aspect of the model, {\it viz}, the statistics of expectation values of certain observables in eigenstates of the Hamiltonian.
\begin{figure}
\includegraphics{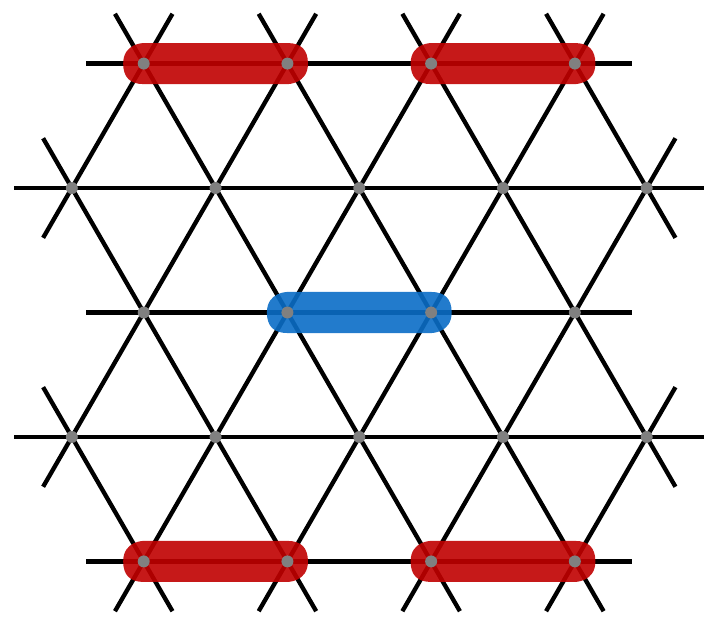}
\caption{Further neighbors used in the definition of the observable \(N_s^\triangle\) for the triangular lattice. For the central link \(\ell\), with a blue dimer, we call the four links \(\ell'\) containing red dimers its ``second neighbors'' \(\goS_\ell\), and define \(N_s^\triangle = \frac{1}{2}\sum_\ell\sum_{\ell'\in\goS_\ell}n_\ell n_{\ell'}\) where \(n_\ell \in \{0,1\}\) is the dimer occupation for link \(\ell\). (For other orientations of the central link \(\ell\), the set \(\goS_\ell\) is defined by applying the rotation symmetries of the lattice.)}
\label{FigTriangleFurtherNeighbors}
\end{figure}
\begin{figure}
\includegraphics{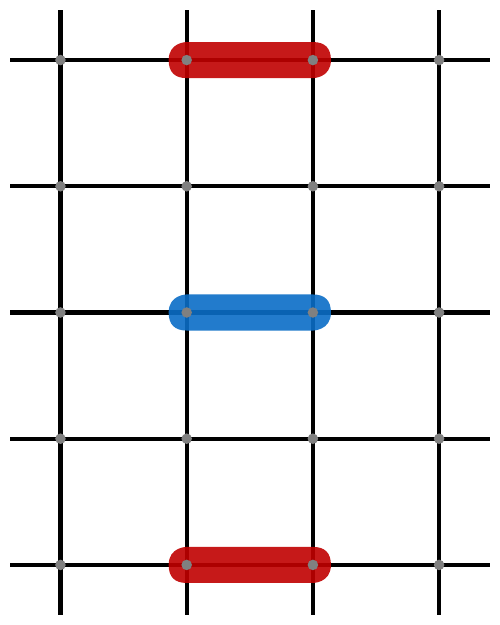}
\caption{Further neighbors used in the definition of the observable \(N_s^\square\) for the square lattice (see \reffig{FigTriangleFurtherNeighbors}). The two red dimers are the ``second neighbors'' of the central blue dimer.}
\label{FigSquareFurtherNeighbors}
\end{figure}
The observables that we choose are: \(N_f^\alpha = \sum_p P_p^\alpha\), i.e., the number of nearest-neighbour parallel dimers; and \(N_s^\alpha\), the number of second-neighbour parallel dimers along the same directions, as illustrated in \reffigand{FigTriangleFurtherNeighbors}{FigSquareFurtherNeighbors}. (The latter definition is chosen because \(N_s^\alpha\) is maximized by the columnar state at \(V = -\infty\), and so \(\langle N_s^\alpha\rangle_T\) is expected to correlate fairly well with \(E_T\).) Both are sums of local operators, and so are expected to obey ETH \cite{FootnoteDiagonal}.

\subsection{Lattice geometries and topological sectors of the QDM}

Our goal is to verify ETH using exact diagonalization of the QDM Hamiltonian. This is made particularly challenging by the fact that ETH is a statistical claim about the approach to the thermodynamic limit, whereas we have a clean system (and so limited statistics) and are necessarily restricted to small system sizes.

The memory requirements for full diagonalization of the Hamiltonian limit us to matrices with linear size below about $5\times10^4$. We show in \reftaband{table_sqr}{table_trig} how the Hilbert-space dimensions for the square and triangular lattices change with system size \cite{DimerNumber1, DimerNumber2, DimerNumber} under periodic boundary conditions (PBCs). (We restrict to clusters of  type A, using the notation of \refcite{Triangular_QDM1}.)
\begin{table}
\caption{ Number of dimer configurations on square lattice with size $L_x \times L_y$ under periodic boundary conditions~\cite{DimerNumber}. Shaded entries show the cases used here.}
\centering
\begin{tabular}{c|c|c|c|c}
\hline \hline
 & $L_y=2$ & $L_y=4$ & $L_y=6$ & $L_y=8$  \\ [0.5ex]
\hline
$L_x=2$  &8 &36&200&1156 \\ \hline
$L_x=3$ &14&50&224&1058 \\ \hline
$L_x=4$ & 36&272&3108&39952\\ \hline
$L_x=5$ & 82&722&9922&155682\\ \hline
$L_x=6$ & 200&\cellcolor{lightgray} 3108& \cellcolor{lightgray}90176&\cellcolor{lightgray}3113860\\ \hline
$L_x=7$ & 478&10082&401998&19681538\\  \hline
$L_x=8$  & 1156&39952&3113860&311853312\\ [1ex]  
\hline \hline
\end{tabular}
\label{table_sqr}
\end{table}
\begin{table}
\caption{Number of dimer configurations on triangular lattice with size $L_x \times L_y$ under periodic boundary conditions~\cite{DimerNumber}. Shaded entries show the cases used here.}
\centering
\begin{tabular}{c|c|c|c|c}
\hline \hline
 & $L_y=2$ & $L_y=4$ & $L_y=6$ & $L_y=8$  \\ [0.5ex]
\hline
$L_x=2$  &12 &72&480&3360 \\ \hline
$L_x=3$ &28&344&4480&58592 \\ \hline
$L_x=4$ & 72&1920&59040&1826944\\ \hline
$L_x=5$ & 184&\cellcolor{lightgray}10608&\cellcolor{lightgray}767776&55801792\\ \hline
$L_x=6$ & 480&59040&10045824&1720316544\\ \hline
$L_x=7$ & 1264&\cellcolor{lightgray}328224&131456320&53046806656\\  \hline
$L_x=8$  & 3360&1826944&1720316544&1635885514752\\ [1ex]  
\hline \hline
\end{tabular}
\label{table_trig}
\end{table}
Exploiting translation symmetry, the largest size of the square lattice we can access is $L_x\times L_y=6\times 8$ with a Hilbert space of dimension $H_D=3113860$, where we diagonalize the Hamiltonian in each momentum sector $(k_x,k_y)$ separately. Similarly, the largest size of the triangular lattice we can access is $L_x\times L_y=4\times 8$ with a Hilbert space of dimension $H_D=1826944$.  Note that the largest momentum sector of $L_x\times L_y=6\times 6$ triangular lattice has a dimension of size 69996, which is beyond our limit.  A further constraint on the lattice sizes we can choose is from the interplay between the shape of the lattice and the topological sectors, which we will discuss next. 

As stated in \refsec{SecIntroduction}, the Hilbert space of the QDM on the square and triangular lattices splits into disconnected sectors due to the nonergodicity of the plaquette-flip dynamics. In three dimensions and higher, there are additional conserved quantities and the Hilbert space within each sector is not connected by single-plaquette flips \cite{3Ddimer3, 3Ddimer4, 3Ddimer5}, but for the square and triangular lattices, all states within each winding-number sector are connected by the kinetic operator, with the exception of some sectors near maximal winding number, which are completely isolated \cite{QDM_review}.

\begin{figure}
\includegraphics[width=1.0\columnwidth]{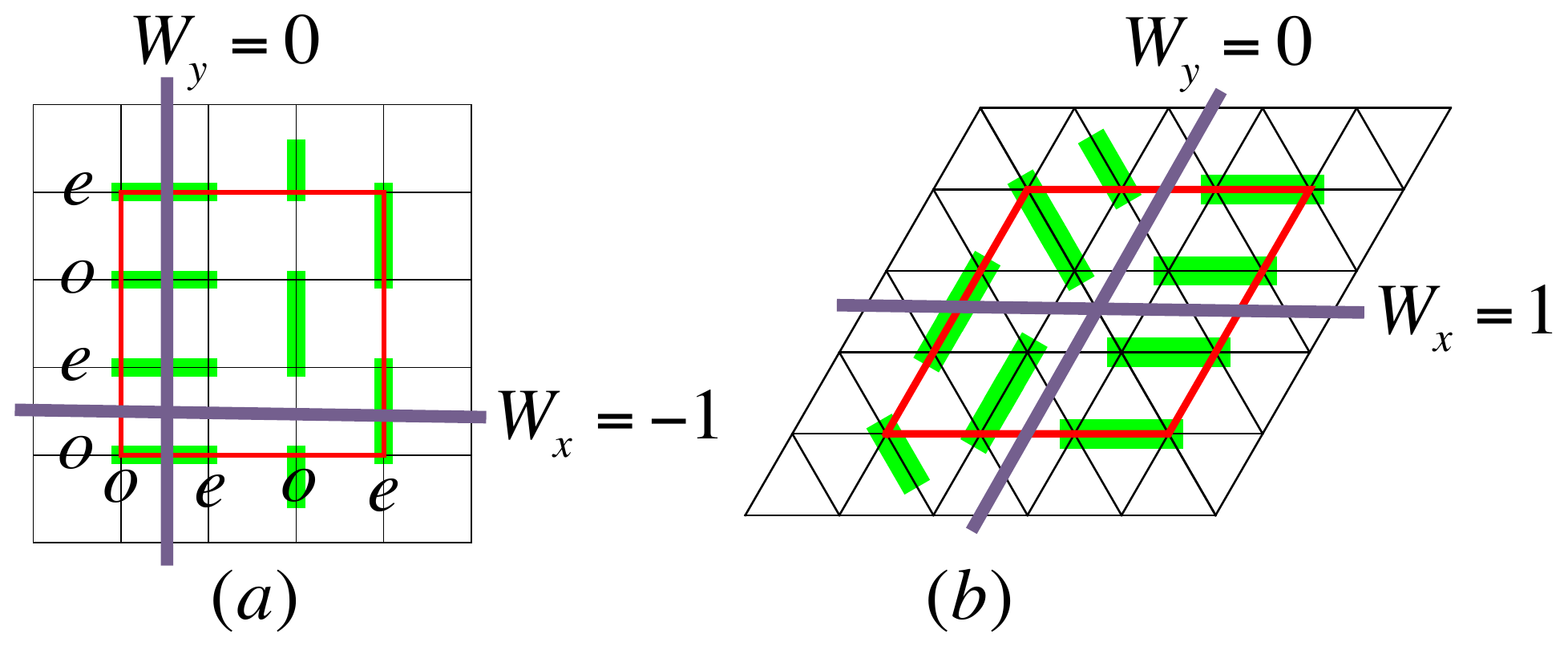}
\caption{Definition of the winding number on square and triangular lattices. For the square lattice, the winding number in the \(x\) direction is $W_x=N\sub{o}-N\sub{e}$, where \(N\sub{o}\) and \(N\sub{e}\) are the number of dimers on links with odd and even \(x\), respectively, that are crossed by the horizontal reference line (and similarly for \(W_y\)). For the triangular lattice, the winding number in direction \(\mu\in\{x,y\}\) is defined by $W_\mu=N \bmod 2$, i.e., the number of dimers crossing the reference line up to its parity. Example dimer configurations and their winding numbers are shown.}
\label{FigWindNum}
\end{figure}

The different sectors can be characterized by topological winding numbers. For a 2D lattice with periodic boundary conditions, i.e., a torus geometry, we can draw reference loops around the \(x\) and \(y\) directions, as illustrated in \reffig{FigWindNum}.
For the square lattice, we label the vertical links that are crossed by the \(x\) loop according to the parity (odd or even) of their \(x\) coordinate, and similarly label the horizontal links crossed by the \(y\) loop by their \(y\) coordinate. For \(\mu \in\{x,y\}\), the winding number \(W_\mu\) is given by \(W_\mu = N\sub{o}-N\sub{e}\), where \(N\sub{o}\) and \(N\sub{e}\) are the number of dimers on links that are crossed by the loop and that are odd and even according to this labeling. Since the plaquette-flip dynamics can only ever add or remove a pair of nearest-neighbour dimers crossing the reference line, it conserves \(W_\mu\). (The same is in fact true of any local dynamics.) The dimer configurations on the square lattice can therefore be split into sectors with given winding numbers $(W_x, W_y)$, where $-L_x/2\leq W_x \leq L_x/2$ and $-L_y/2\leq W_y\leq L_y/2$.

Because the triangular lattice is not bipartite, there is no consistent way to label links crossed by the reference loop as odd or even. Instead the winding number is defined by the number of dimers crossing the reference line up to parity, \(W_\mu = N \bmod 2\). There are therefore only four sectors, with $(W_x, W_y)=(0,0)$, $(0,1)$, $(1,0)$, and $(1,1)$. Because every plaquette is crossed by any chosen reference line an even number of times, these winding numbers are also conserved by the dynamics (and by any local dynamics).

We will show that the manifestation of ETH in the QDM is sensitive to how these topological sectors are treated. Since different system geometries imply different degeneracies between the sectors, it is helpful to choose sizes with consistent degeneracy structures. For the square lattice, it is sufficient to restrict to the case where $L_x$ and $L_y$ are both even. We therefore take \((L_x, L_y)=(4,6)\), \((6,6)\), \((6,8)\), the three largest accessible system sizes.

This issue is more significant for the triangular lattice, where we must account for the fact that spatial symmetry operations change the winding numbers and hence map configurations from one sector to another. This implies that energy eigenstates in different sectors are related by symmetry operations, and have identical energies as well as identical expectation values of symmetric observables such as \(N_f^\triangle\) and \(N_s^\triangle\).

The effect of the symmetries on the winding numbers, and hence on the degeneracies, is determined by the shape of the system, and in particular on its size in units of the lattice vectors. (We consider only clusters of type A, according to the classification of \refcite{Triangular_QDM1}.) To host close-packed dimer configurations, the lattice must have an even number of sites, and so there are three distinct cases:
\begin{itemize}
\item even $L_x = L_y$---Rotations by \(\pi/3\) relate configurations in three of the sectors \cite{Triangular_QDM1}. The largest system size accessible to full diagonalization is \(L_x = L_y = 4\).

\item $L_x \neq L_y$, both even---In this case, the four sectors are all different. The largest accessible sizes are $(L_x,L_y)=(4,6)$ and $(4,8)$.

\item \(L_x\) even, \(L_y\) odd---In this case, symmetries relate the \((0,0)\) and \((0,1)\) sectors as well as the \((1,0)\) and \((1,1)\) sectors \cite{FootnoteTranslationSymmetry}. Accessible sizes include $(L_x,L_y)=(4,5)$, $(4,7)$, and $(6,5)$.
\end{itemize}
We therefore choose \(L_x\) even and \(L_y\) odd, so that we have three accessible system sizes with the same degeneracy structure. A negative trade-off of this choice is that we must include the case \((4,7)\), which has relatively large anisotropy.


\section{Results}
\label{SecResults}

In this section we will present our main results, first for the square lattice and then the triangular lattice. The primary quantities of interest are the expectation values of the observables \(N^\alpha_f\) and \(N^\alpha_s\) in energy eigenstates. For each lattice, we first present the raw data, from which some qualitative features can be understood. We then analyze statistical properties of the expectation values and compare these with predictions based on ETH.

In the following, we will set $t=1$ as the energy unit and vary $V$ only. Only the cases with $V\geq0$ are considered because, as we prove in the \refapp{AppSymmetry}, the spectra and diagonal matrix elements are identical for $V$ and $-V$.

\subsection{ETH for the square QDM}

\begin{figure*}
\centering
\includegraphics[width=\textwidth]{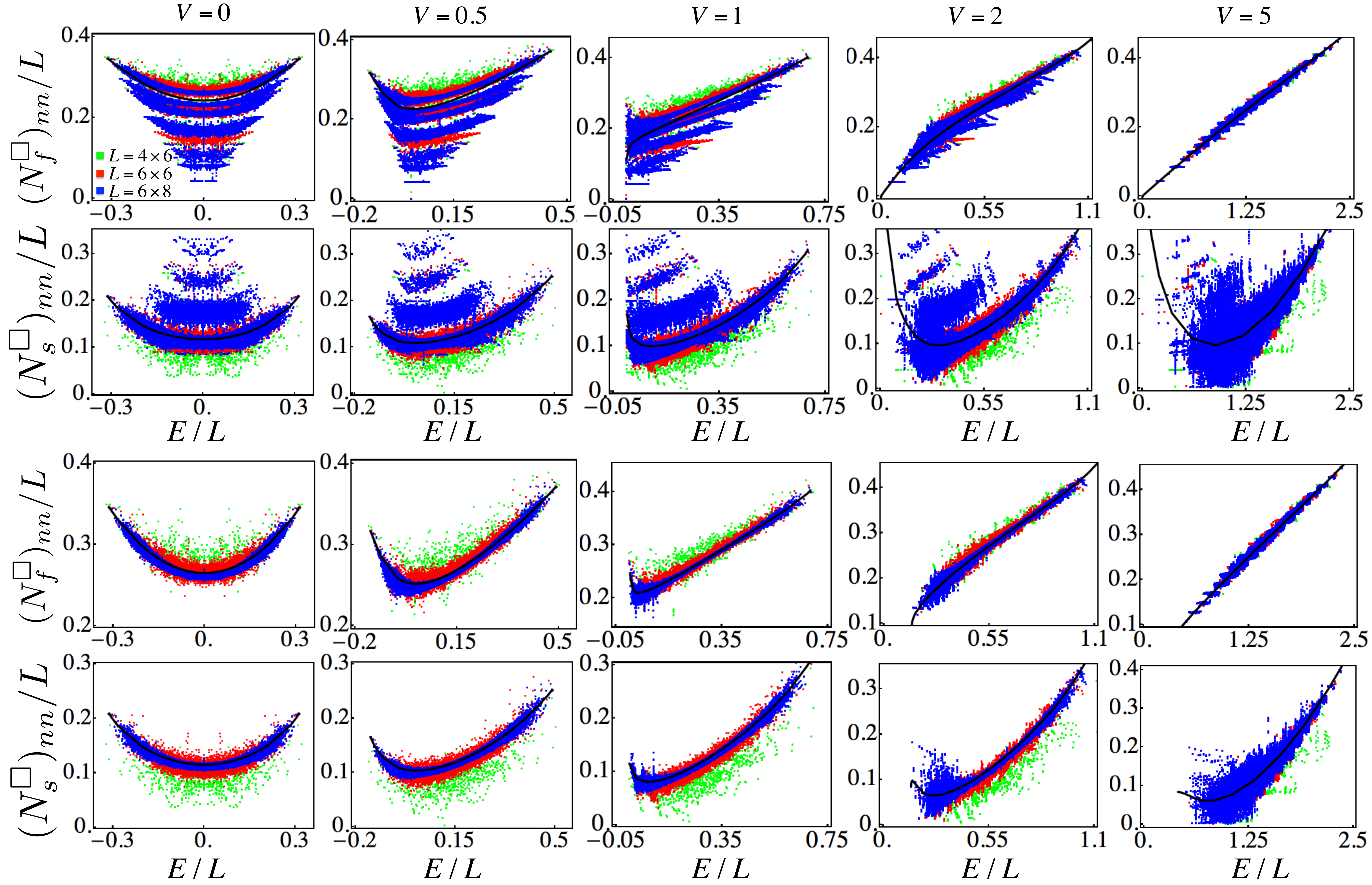}
\caption{Diagonal matrix elements of $N_f$ and $N_s$ in the energy eigenbasis for the QDM on square lattices with size $L=L_x \times L_y=4\times6$, $6\times6$ and $6\times8$ at $V=0, 0.5, 1, 2, 5$. The top two rows show the full spectrum, containing all the topological sectors, while the bottom two rows show the spectrum from only the \((0,0)\) topological sector. The solid black lines show the expectation values in the canonical ensemble, given by \refeq{EqThermalEV}, at a temperature \(T\) chosen such that the mean energy is \(E\) (for the largest lattice size, $L=6 \times 8$).}
\label{f1}
\end{figure*}
In \reffig{f1}, we show \((N_f^\square)_{nn}\) and \((N_s^\square)_{nn}\), the diagonal matrix elements of $N_f$ and $N_s$, in all eigenstates $\lvert E_n\rangle$ of the Hamiltonian \(H^\square\) for the square lattice, at various values of $V$ (columns) and lattice sizes $L = L_xL_y$ (colors). The top two rows include the full spectrum, containing all topological sectors, while the bottom two rows restrict to the \((0,0)\) sector.

It is interesting to see that, for small \(V\), the full spectra of both $N_f$ and $N_s$ (top two rows of \reffig{f1}) form a structure consisting of multiple bands. We have verified that this results from the fact that each topological sector forms a band and some of these overlap. Concentrating on the \((0,0)\) sector (bottom two rows of \reffig{f1}), which contains roughly half of the whole Hilbert space, we see that the distribution becomes narrower as the system size increases, at least in the middle of the distribution and for \(V\) not too large. Furthermore, the expectation values in eigenstates of energy \(E\) appear to converge towards the thermal expectation value (black line) at a temperature chosen such that the mean energy is \(E\), consistent with the predictions of ETH as detailed in \refsec{SecETH}.

Finally, for large $V$, the QDM approaches a trivially integrable point at $V=\infty$, where the Hamiltonian contains only the diagonal terms $P_p^\alpha$ and thus reduces to a classical Hamiltonian. This is reflected in the step structure in the case \(V = 5\), which is already visible within the \((0,0)\) sector.

To provide a more quantitative picture, we employ an approach that has proved useful in previous studies \cite{2D_ETH, constrained_anyon}, by looking at fluctuations between the diagonal matrix elements in adjacent energy eigenstates. We first sort all the eigenstates by energy and then calculate the difference of diagonal matrix elements between adjacent eigenstates, 
\begin{equation}
(\Delta N_{f,s})_{n}=(N_{f,s})_{n+1,n+1}-(N_{f,s})_{n,n}\punc{.}
\label{rms}
\end{equation}
As discussed in \refsec{SecETH}, ETH predicts that the width of the distribution of \(\Delta N_{f,s}\) should decrease exponentially with system size. We note that since it is well known that states at the edges of the spectrum do not exhibit eigenstate thermalization (as can also be seen clearly from the bottom two rows of \reffig{f1}), we will only keep the middle part of the spectrum. We follow \refcite{2D_ETH} and retain only states with energy $E_n$ such that
\begin{equation}
\label{EqThreshold}
 \lvert E_n-E\sub{mid} \rvert < x\sub{thr} \frac{E\sub{max}-E\sub{min}}{2}
 \end{equation}
where \(E\sub{min}\) and \(E\sub{max}\) are the eigenstates of minimal and maximal energy respectively, $E\sub{mid}=(E\sub{max}+E\sub{min})/2$ and $x\sub{thr}$ is the truncation parameter, which we will set to $x\sub{thr}=0.2$.

\begin{figure*}
\centering
\includegraphics[width=2.0\columnwidth]{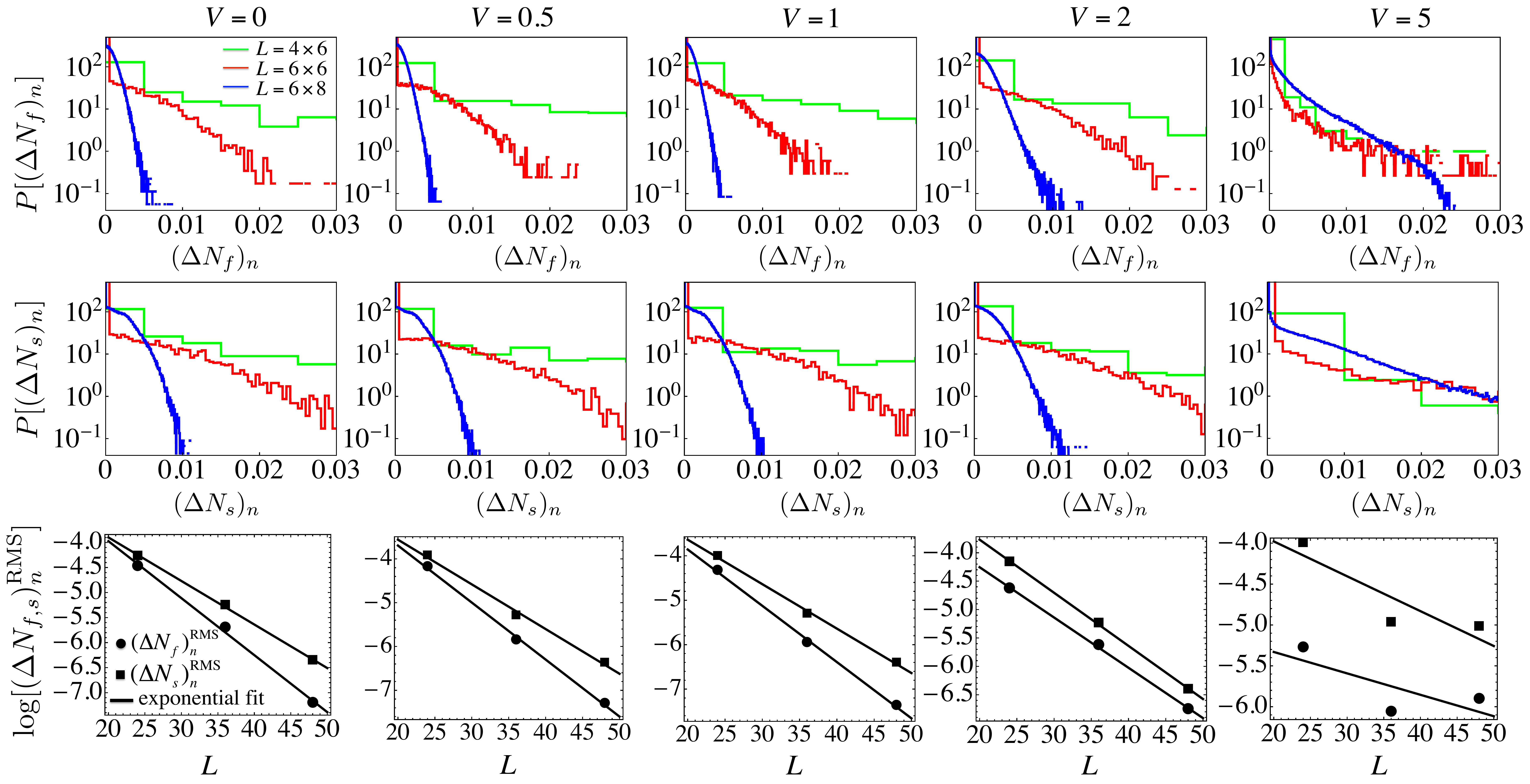}
\caption{Top two rows: Distribution of $(\Delta N_f)_n$ and $(\Delta N_s)_n$, the difference of the diagonal matrix element of \(N_f\) and \(N_s\) for successive energy eigenstates, on the square lattice. Only the \((0,0)\) topological sector, corresponding to the bottom two rows of \reffig{f1}, is included. States at the edges of the spectrum, i.e., those not obeying \refeq{EqThreshold} with $x\sub{thr}=0.2$ for all the cases, are excluded. Bottom row: Root-mean-square (RMS) values of $(\Delta N_f)_n$ and $(\Delta N_s)_n$. The solid lines show an exponential fit (note the logarithmic scale on the vertical axis), to test the ETH claim that the eigenstate-to-eigenstate fluctuations of the diagonal matrix elements are suppressed exponentially with system size.}
\label{f2}
\end{figure*}
The top two rows of \reffig{f2} show the distributions of $(\Delta N_f)_n$ and $(\Delta N_s)_n$, which clearly narrow as \(L\) increases, for \(V \le 2\). Also shown, in the bottom row of \reffig{f2}, is the root-mean-square (RMS) of these distributions, calculated after the truncation. From the exponential fit of the RMS values of $(\Delta N_f)_n$ and $(\Delta N_s)_n$ (note the logarithmic scale on the vertical axis of the bottom row of \reffig{f2}), we see the hallmark of ETH, i.e., that the eigenstate-to-eigenstate fluctuations of the diagonal matrix elements of the observables decrease exponentially with system size, is evident for the smaller values $V$. At $V=5$, ETH appears to break down, with no consistent decrease in the RMS with increasing system size.

\subsection{ETH for the triangular QDM}

\begin{figure*}
\centering
\includegraphics[width=\textwidth]{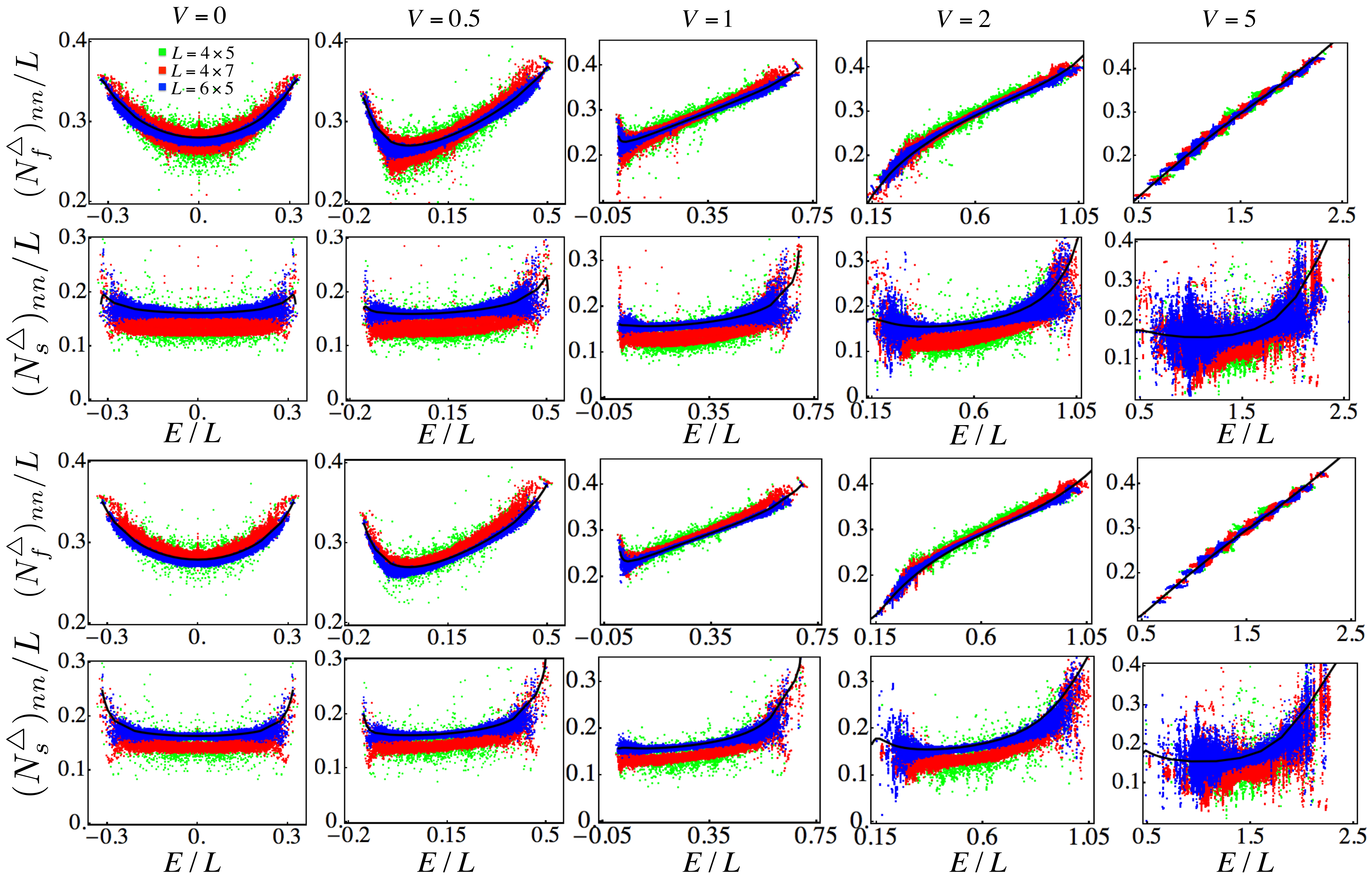}
\caption{Diagonal matrix elements of $N_f$ and $N_s$ in the energy eigenbasis for the QDM on triangular lattices with size $L=L_x \times L_y=4\times5$, $4\times7$ and $6\times5$ at $V=0, 0.5, 1, 2, 5$. The top two rows show the full spectrum, containing all the topological sectors, while the bottom two rows show the spectrum from only the \((0,0)\) topological sector. The solid black lines show the expectation values in the canonical ensemble, given by \refeq{EqThermalEV}, at a temperature \(T\) chosen such that the mean energy is \(E\) (for the largest lattice size, $L=6\times 5$).}
\label{f3}
\end{figure*}
In \reffig{f3}, we show the raw data for the triangular lattice. The diagonal matrix elements of $N_f$ and $N_s$ are shown for all topological sectors in the top two rows, and within the \((0,0)\) sector in the bottom two rows. (Recall that there are four sectors for the triangular lattice, but, for the system sizes used here, symmetries reduce these to two distinct pairs.) In this case, the distribution becomes narrower with increasing system size \(L\) both for the full spectrum and when restricted to a single sector.

\begin{figure*}
\centering
\includegraphics[width=2.0\columnwidth]{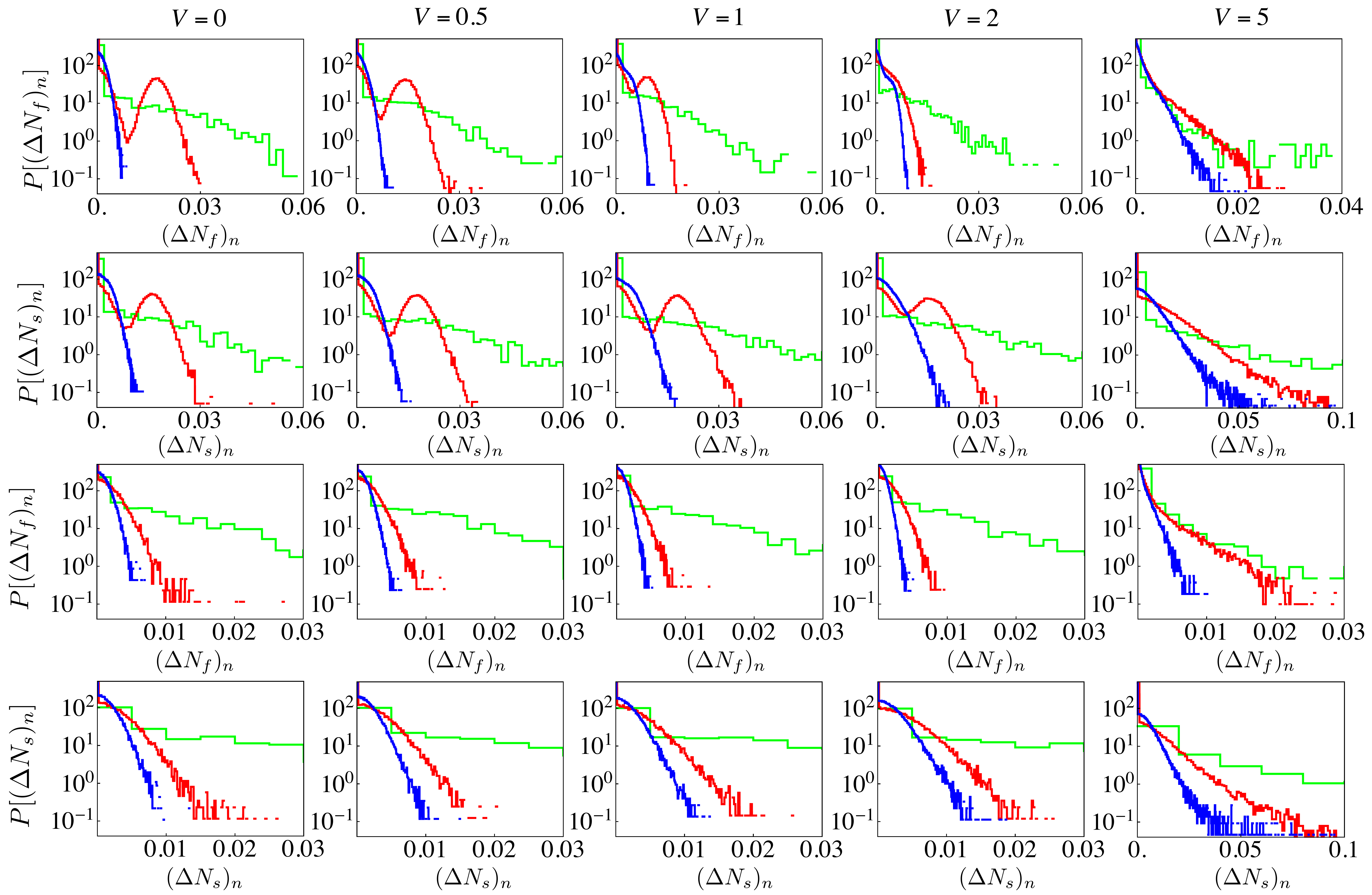}
\caption{Distribution of $(\Delta N_f)_n$ and $(\Delta N_s)_n$, the difference of the diagonal matrix element of \(N_f\) and \(N_s\) for successive energy eigenstates, on the triangular lattice. The top two rows show the full spectrum, containing all the topological sectors, while the bottom two rows show the spectrum from only the \((0,0)\) topological sector. To exclude states at the edges of the spectrum, which do not exhibit ETH, only states obeying \refeq{EqThreshold} are included, with $x\sub{thr}=0.2$ in all cases.}
\label{f4}
\end{figure*}
To determine whether this narrowing is quantitatively consistent with ETH, we calculate $(\Delta N_f)_n$ and $(\Delta N_s)_n$, defined by \refeq{rms}, for the triangular lattice. In \reffig{f4}, we show the distribution of these quantities, calculated after truncation using \refeq{EqThreshold} with $x\sub{thr}=0.2$. As was already evident from \reffig{f3}, the distributions become narrower with increasing $L$ for both the full spectrum and that of the \((0,0)\) sector.

For $L=4\times7$, there is a clear two-peak structure that is visible for the spectrum treated as a whole but not for the \((0,0)\) sector. This indicates a separation of the expectation values into bands corresponding to different topological sectors, as for the square lattice. In this case, however, there is no visible splitting for the more isotropic systems, $L=4\times5$ and $6\times5$.

\begin{figure*}
\centering
\includegraphics[width=\textwidth]{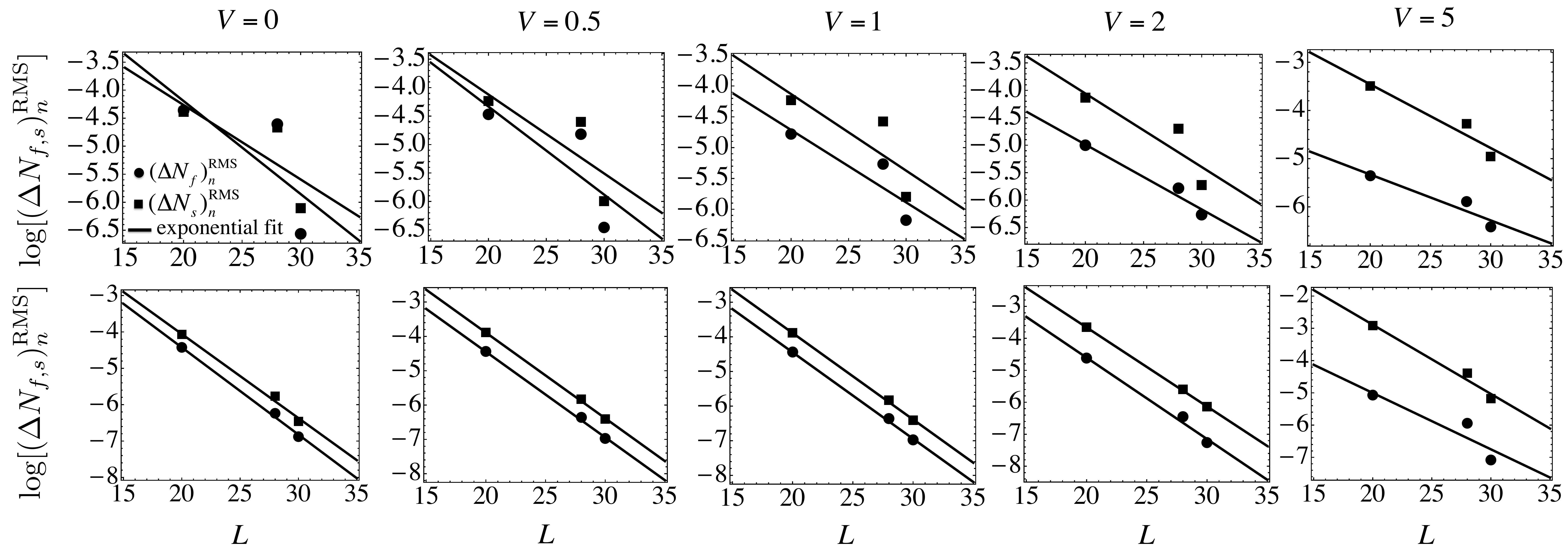}
\caption{Root-mean-square (RMS) values of $(\Delta N_f)_n$ and $(\Delta N_s)_n$ for the triangular lattice. The solid lines show an exponential fit (note the logarithmic scale on the vertical axis), to test the ETH claim that the eigenstate-to-eigenstate fluctuations of the diagonal matrix elements are suppressed exponentially with system size. The top row shows the results from the full spectrum, containing all the topological sectors, while the bottom row shows the results only from the \((0,0)\) topological sector.}
\label{f5}
\end{figure*}
In \reffig{f5}, we present the RMS values of $(\Delta N_f)_n$ and $(\Delta N_s)_n$. As for the square lattice, a reasonable fit is achieved to an exponential decrease with system size, but only within the \((0,0)\) sector. It should be noted that the more isotropic systems, $L=4\times5$ and $6\times5$, show a consistent decrease, similar to that for a single sector, even with the spectrum treated as a whole. One might speculate that the breakdown of the ETH in this case is due to the confounding influence of the variation in system geometry.


\section{Conclusions and outlook}
\label{SecConclusions}

We have studied the statistical properties of energy-eigenstate expectation values in the quantum dimer model (QDM) on both the square and triangular lattices, as a test of the applicability of the eigenstate thermalization hypothesis (ETH). We find results consistent with the quantitative predictions of ETH, as long as one treats separately the different topological sectors of the Hilbert space, which are disconnected under the action of the Hamiltonian. When the spectrum is treated as a whole, we observe substantial deviations from ETH, for the system sizes that are accessible in exact diagonalization.

While the evidence for the breakdown of ETH as applied to the full Hilbert space is, we believe, convincing for the square lattice, the situation is less clear for the triangular lattice. In the latter case, the set of system sizes accessible by exact diagonalization is extremely limited, and the choice is complicated by the interaction between system geometry and topological degeneracies. The three systems for which we have presented results have a range of aspect ratios, introducing an additional confounding variable that may be obscuring the exponential trend expected from ETH.

From a theoretical point of view, the distinction between square and triangular lattices results from their different topological conservation laws, which are in turn due to the fact that the triangular lattice is not bipartite. The winding numbers \(W_x\) and \(W_y\) can therefore be defined only up to parity, and there are only four sectors, some of which are related by symmetries. It is possible that the different sectors may become indistinguishable in the thermodynamic limit; more precisely, full restoration of ETH would require that the sector-dependent effects become exponentially small in the system size \(L\).

The dimer model on the bipartite square lattice, by constrast, has integer winding numbers and thus an extensive number of topological sectors. It is therefore reasonable to expect a spread in their properties that remains in the thermodynamic limit. It should be noted, though, that for large system sizes, most states (in fact, all but an exponentially small fraction) have winding numbers that are much less than the maximal value, of the order of the linear system size. In this respect, the square lattice becomes, for larger \(L\), more like the triangular lattice, where all states have \(W_\mu\) of the order of unity.

For the square lattice, the possibility therefore remains that, in the thermodynamic limit, the subset of states with \(W_\mu\) of the order of unity obeys ETH when treated as a whole. As for the triangular lattice, this would require that their dependence on \(W_\mu\) not merely decreases, but does so exponentially with \(L\). With the system sizes available, it is not possible to state definitively whether this will indeed happen. In any case, it is certainly true that the minority of eigenstates that have extensive winding numbers will remain outliers as far as ETH is concerned, for any finite size.

While the need to distinguish topological sectors may not be viewed as particularly surprising, it should be noted that this is apparently unnecessary with other global conserved quantities, such as momentum. (We have exploited momentum conservation to aid diagonalization of the Hamiltonian, but all momentum sectors are included in our results.) The winding numbers defining the topological sectors are apparently more similar in this respect to local conserved quantities in integrable models \cite{integrable_review}. In principle it should be possible to include all topological sectors on an equal footing, by analogy with the GGE for integrable systems \cite{GGE}, but it remains to be seen whether this can be done in practice.

In this work, we have studied the predictions of ETH only for diagonal matrix elements of the observables. Predictions for off-diagonal elements are also contained in Eq. (1), but an alternative perspective on this question involves the time dependence of observables after a quantum quench \cite{review2}. Our results here suggest that interesting behavior should occur at large $V$ (around $V/t \gtrsim 5$), where ETH begins to break down, and we indeed observe signatures of slow relaxation and metastability at such parameters \cite{dimer_dyn}. Another interesting extension is the disordered QDM and its possible connections to many-body localization \cite{MBL}.

\section{Acknowledgments}

The simulations used resources provided by the University of Nottingham High-Performance Computing Service. We are grateful to F. Alet, J. P. Garrahan, and M. Marcuzzi for helpful discussions. This work was supported by EPSRC Grant No.~EP/M019691/1. 

{\it Data availability}. Research data are available from the Nottingham Research Data Management Repository at http://dx.doi.org/10.17639/nott.327.

\appendix 

\section{Symmetry between \(+V\) and \(-V\)}
\label{AppSymmetry}

In this appendix, we prove that there exists a unitary operator \(\scR\) such that, for any eigenstate \(\ket{E}\) of $H^\alpha(V)$ with eigenvalue \(E\), \(\scR\ket{E}\) is an eigenstate of $H^\alpha(-V)$ with eigenvalue \(-E\). Furthermore, any operator that is diagonal in the basis of the dimer configurations, such as \(N_{f}^\alpha\) and \(N_{s}^\alpha\), commutes with \(\scR\) and hence has the same diagonal matrix element in \(\ket{E}\) and \(\scR\ket{E}\).

\begin{figure}
\centering
\includegraphics[width=1.0\columnwidth]{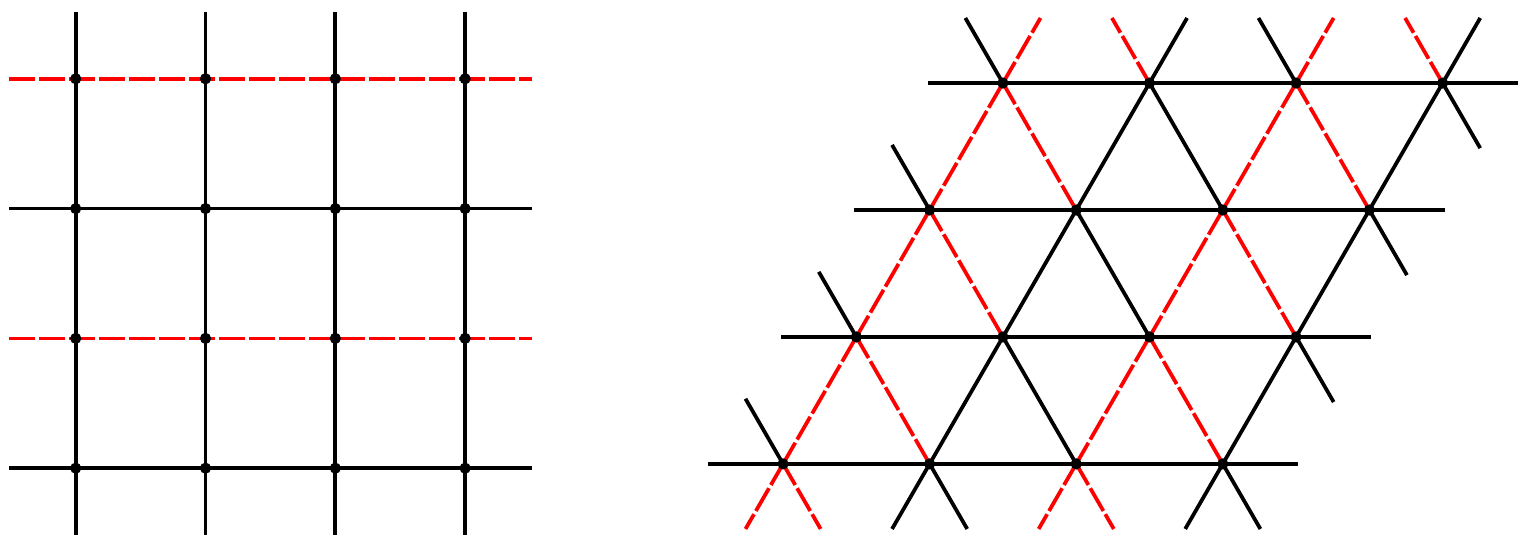}
\caption{Pattern of bonds used in the definition of the operator \(\scR\) (see \refapp{AppSymmetry}). Each plaquette (square for the square lattice, left; rhombus for the triangular lattice, right) has an odd number of red (dashed) links.}
\label{s1}
\end{figure}
For any dimer configuration $\Psi$, let $n$ be the number of dimers occupying
links colored red in \reffig{s1}, and let $R(\Psi)=(-1)^n$. Since every
plaquette has an odd number of red links, flipping any plaquette
gives a configuration $\Psi'$ with $R(\Psi')=-R(\Psi)$. We can define the unitary (in fact involutory) operator $\scR$,
which is diagonal in the basis of dimer configurations $\ket{\Psi}$ and has the property $\scR \ket{\Psi}=(-1)^n \ket{\Psi}$.

Now since $\scR$ and $P_p^\alpha$, defined in \refsec{SecModels}, are both diagonal in the basis $|\Psi\rangle$, they commute, i.e., $\mathcal{R}P_p^\alpha = P_p^\alpha\scR$. On the other hand, because each term of $T_p^\alpha$ flips a plaquette, which changes the sign of \(R\), the operators $T_p^\alpha$ and $\scR$ anticommute, i.e., $\scR T_p^\alpha=-T_p^\alpha\scR$. Applying the unitary transformation $\scR$ to the Hamiltonian, given by \refeq{eqn:QDMs}, therefore gives
\begin{multline}
\scR H^\alpha(V) \scR^{-1} = \sum_p \left(-t \scR T_p^\alpha \scR^{-1} + V \scR P_p^\alpha \scR^{-1} \right)\\
=\sum_p\big(t T_p^\alpha+ V P_p^\alpha\big) =-H^\alpha(-V)\punc{.}
\end{multline}
The statement that \(\scR\ket{E}\) is an eigenstate of \(H^\alpha(-V)\) with eigenvalue \(-E\) follows. Finally, \(\scR\) is by definition diagonal in the basis of dimer configurations, and hence commutes with any other such operator.



\begin{thebibliography}{99}

\newcommand{\journal}[4]{#1 {\bf #2}, #3 (#4)}
\newcommand{\journaltitle}[5]{{\it #1}, \journal{#2}{#3}{#4}{#5}}
\newcommand{\journaltitlenovolume}[4]{{\it #1}, #2 ({\bf #4}), #3}
\newcommand{\book}[3]{{\it #1} (#2, #3)}
\newcommand{\prx}{Phys. Rev. X}
\newcommand{\arcmp}{Annu. Rev. Cond. Matt. Phys.}
\newcommand{\PR}[3]{\journal{\pr}{#1}{#2}{#3}}
\newcommand{\PRA}[3]{\journal{\pra}{#1}{#2}{#3}}
\newcommand{\PRB}[3]{\journal{\prb}{#1}{#2}{#3}}
\newcommand{\PRL}[3]{\journal{\prl}{#1}{#2}{#3}}
\newcommand{\PRX}[3]{\journal{\prx}{#1}{#2}{#3}}

\bibitem{review1} C. Gogolin and J. Eisert,  Rep. Prog. Phys. {\bf 79}, 056001 (2016). 
\bibitem{review2} L. D. Alessio, Y. Kafri, A. Polkovnikov, and M. Rigol, Adv. Phys. {\bf 65}, 239 (2016). 
\bibitem{review3} F. Borgonovi, F.M. Izrailev, L.F. Santos and V.G. Zelevinsky, Phys. Rep. {\bf 626}, 1 (2016). 
\bibitem{Srednicki1999} M. Srednicki, J. Phys. A {\bf 32}, 1163 (1999).


\bibitem{integrable_review} P. Calabrese, F. H. L Essler and G. Mussardo,  J. Stat. Mech. Theor. Exp. (2016) 064001.
\bibitem{exp1}T. Langen, S. Erne, R. Geiger, B. Rauer, T. Schweigler, M. Kuhnert, W. Rohringer, I. E. Mazets, T. Gasenzer and J. Schmiedmayer, Science {\bf 348}, 207 (2015). 
\bibitem{exp2} A. M. Kaufman, M. E. Tai, A. Lukin, M. Rispoli, R. Schittko, P. M. Preiss and M. Greiner, Science {\bf 353}, 794 (2016).


\bibitem{QDM_review} R. Moessner and K. S. Raman, in {\it Introduction to Frustrated Magnetism: Materials, Experiments, Theory}, edited by C. Lacroix, F. Mila and P. Mendels (Springer, Berlin, 2011).
\bibitem{QDM_RK} D. S. Rokhsar and S. A. Kivelson, Phys. Rev. Lett. {\bf 61}, 2376 (1988).

\bibitem{Rydberg_QDM} A. W. Glaetzle, M. Dalmonte, R. Nath, I. Rousochatzakis, R. Moessner and P. Zoller, Phys. Rev. X 4, 041037 (2014). 
\bibitem{cold_QDM} B. Sundar, T. C. Rutkowski, E. J. Mueller and M. J. Lawler, arXiv:1702.05514.

\bibitem{QDM_liquid1} R. Moessner and S. L. Sondhi, Phys. Rev. Lett. {\bf 86}, 1881 (2001).
\bibitem{QDM_liquid2} G. Misguich, D. Serban, and V. Pasquier, Phys. Rev. Lett. {\bf 89}, 137202 (2002).



\bibitem{2D_ETH} R. Mondaini, K. R. Fratus, M. Srednicki, and M. Rigol, Phys. Rev. E {\bf 93}, 032104 (2016). 
\bibitem{2D_ETH_off} R. Mondaini and M. Rigol, Phys. Rev. E {\bf 96}, 012157 (2017).
\bibitem{2D_TFIM} B. Bla\ss  ~and H. Rieger, Sci. Rep.  {\bf 6} 38185 (2016). 

\bibitem{constrained_anyon}A. Chandran, M. D. Schulz, and F. J. Burnell, Phys. Rev. B {\bf 94}, 235122 (2016). 
\bibitem{UedaPRE16} R. Hamazaki, T. N. Ikeda, and M. Ueda, Phys. Rev. E {\bf 93}, 032116 (2016). 

\bibitem{1D_ETH1}M. Rigol, Phys. Rev. Lett. {\bf 103}, 100403 (2009).
\bibitem{1D_ETH2} T. N. Ikeda, Y. Watanabe, and M. Ueda, Phys. Rev. E {\bf 87}, 012125 (2013).
\bibitem{1D_ETH3} S. Sorg, L. Vidmar, L. Pollet, and F. Heidrich-Meisner, Phys. Rev. A {\bf 90}, 033606 (2014). 
\bibitem{1D_ETH4} V. Alba, Phys. Rev. B {\bf 91}, 155123 (2015).

\bibitem{Square_QDM} S. Sachdev, Phys. Rev. B {\bf 40}, 5204 (1989).
\bibitem{Triangular_QDM1} A. Ralko, M. Ferrero, F.  Becca, D. Ivanov, and F. Mila, Phys. Rev. B {\bf 71}, 224109 (2005). 
\bibitem{Triangular_QDM2} H. Ribeiro, S. Bieri, and D. Ivanov, Phys. Rev. B {\bf 76}, 172301 (2007).
\bibitem{Topo_blocking} G. Kells, D. Sen, J. K. Slingerland, and S. Vishveshwara, Phys. Rev. B {\bf 89}, 235130 (2014).

\bibitem{FootnoteDiagonal}{Both \(N_f^\alpha\) and \(N_s^\alpha\) are diagonal in the basis of dimer configurations. The only observables that are off-diagonal but consistent with the dimer constraints are those that shift dimers around closed loops. The simplest such operator is the plaquette-flip operator \(T_p^\alpha\), but summing that over all plaquettes gives the kinetic energy; within an eigenstate of the total energy, looking at this is equivalent to looking at the potential-energy operator \(N_f^\alpha\). One could look at kinetic operators for longer loops, but these are more complicated objects whose expectation values are small in absolute terms, and it is not clear that one gains much from looking at them. We note that there is no particular reason to expect observables that are diagonal in the dimer basis to be either more or less likely to obey ETH.}

\bibitem{DimerNumber1} M. E. Fisher, Phys. Rev. {\bf 124}, 1664 (1961).
\bibitem{DimerNumber2} P. W. Kasteleyn, Physica  {\bf 27}, 1209 (1961).
\bibitem{DimerNumber} N. Sh. Izmailian and R. Kenna, Phys. Rev. E {\bf 84}, 021107 (2011). 

\bibitem{3Ddimer3} F. S. Nogueira  and Z. Nussinov, Phys. Rev. B {\bf 80}, 104413 (2009).
\bibitem{3Ddimer4} O. Sikora, F. Pollmann, N. Shannon, K. Penc and P. Fulde, Phys. Rev. Lett. {\bf 103}, 247001 (2009).
\bibitem{3Ddimer5} O. Sikora, N. Shannon, F. Pollmann, K. Penc and P. Fulde, Phys. Rev. B {\bf  84}, 115129 (2011).


\bibitem{FootnoteTranslationSymmetry}{In fact the relationship is effected by a translation of one lattice site along the \(x\) direction. This implies that the translation symmetry of the lattice is effectively reduced, which must be taken into account when block-diagonalizing the Hamiltonian in momentum space.}



\bibitem{GGE} L. Vidmar and M. Rigol, J. Stat. Mech. Theor. Exp. (2016) 064007.
\bibitem{dimer_dyn} Z. Lan, M. van Horssen, S. Powell, and J. P. Garrahan, arXiv:1706.02603.
\bibitem{MBL} R. Nandkishore and D. A. Huse, Annu. Rev. Condens. Matter Phys. {\bf 6} 15, (2015).

\end{thebibliography}
\end{document}